\documentclass[10pt]{article}
\usepackage{amsmath,amssymb,enumerate} 

\usepackage{amsthm}

\usepackage{tikz}
\usetikzlibrary{arrows}
\usetikzlibrary{positioning}

\def\ZZ{\mathbb{Z}}
\def\ZZ1{\mathbf{Z}_{\geq 1}}

\def\RR{\mathbb{R}}

\title{On Nash-solvability of finite
$n$-person shortest path games; bi-shortest path conjecture  
}

\author{
Vladimir Gurvich\thanks{
HSE University, Moscow Russia; e-mail:
vgurvich@hse.ru ; vladimir.gurvich@gmail.com}
}

\begin{document}

\maketitle

\begin{abstract}
We formulate a conjecture from graph theory that is equivalent 
to Nash-solvability of the finite two-person shortest path games 
with positive local costs. 
For the three-person games such conjecture fails. 
\end{abstract}

\section{Bi-shortest path conjecture} 
\label{s1}
Let    $G = (V,E)$   be a finite directed graph (digraph) 
with two distinct vertices  $s, t \in V$.   
We assume that 
\begin{itemize} 
\item[(j)] every vertex  $v \in V \setminus \{t\}$  has an outgoing edge, 
while  $t$  has not;  
\item[(jj)] $G$ contains a directed path from  $s$ to $t$;  
\item[(jjj)] every edge  $e \in E$  belongs to such a path. 
\end{itemize} 
If (j) fails for  $v$ we merge  $v$ and  $t$; 
if (jjj)  fails for  $e$  we delete  $e$  from  $E$.

\medskip   

Given a partition  $V \setminus \{t\} = V_1 \cup V_2$  
with non-empty  $V_1$  and  $V_2$, 
assign an ordered pair of positive real numbers  $(r_1(e), r_2(e))$  to every $e \in E$.

Fix  $ i \in \{ 1, 2\}$  and  a mapping  $s_i$  that assigns 
to each  $v \in  V_i$  an edge  $e \in E$  going from  $v$.
Delete all other edges going from  $v$. 
In the obtained digraph find a directed shortest path 
(SP)  from  $s$  to  $t$,  assuming that  
$r_{3-i}(e)$  are the lengths of the edges  $e \in E$. 
(One can use, for example, Dijkstra's SP algorithm.)
Doing so for  $i = 1, 2$  and for every  $s_i$  we obtain two sets of 
directed  $(s,t)$-paths. 
We  conjecture that these two sets intersect and 
call this statement the {\em Bi-SP conjecture}. 

\medskip 

Without loss  of  generality 
(WLOG) we can assume that all  $(s,t)$-paths 
have pairwise different  lengths. 

\medskip  

It may happen that some mappings  $s_i$  leave no $(s,t)$-path.  
Then, we choose nothing. 
Let us slightly modify the procedure  
choosing in this case some symbolic path $c$. 
Then we obtain a weak versions of the Bi-SP conjecture.  
Indeed, if the obtained two sets of $(s,t)$-paths have only  $c$ in common    
then the Bi-SP conjecture fails, but the weak Bi-SP one holds. 
   
\medskip

WLOG, we can restrict ourselves by the bipartite graphs 
with parts $(V_1,V_2)$. 
Indeed, if  $E$  contains an edge  $e = (u,w)$   such that both 
$u,w \in V_i$, we subdivide  $e$  by a vertex  $v \in V_{3-i}$ 
into two edges  $e' = (u,v)$  and  $e'' = (v,w)$  choosing some lengths  
$r_i(e') > 0$  and  $r_i(e'') > 0$  such that 
$r_i(e) = r_i(e') + r_i(e'')$  for $i = 1,2$.

\section{Finite $n$-person shortest path games} 
\label{s2}
\subsection*{Players, positions, moves, and local costs}  
Given a finite digraph  $G =(V, E)$  
satisfying assumption (j, jj, jjj) of Section~\ref{s1}, 
let us generalize case  $n=2$ and consider an arbitrary integer  $n \geq 2$.
Partition vertices into  $n$  non-empty subsets 
$V \setminus t = V_1 \cup \ldots \cup V_n$, assign an ordered $n$-tuple 
of positive real numbers  $r(e) = (r_1(e), \ldots, r_n(e))$  to each $e \in E$,  
and consider the following interpretation: 
$I = \{1, \ldots, n\}$  is a set of {\em players}, 
$V_i$  the set of {\em positions} controlled by player $i \in I$; 
furthermore, $s = v_0$  and  $t = v_t$  are respectively 
the {\em initial} and {\em terminal} positions; 
$e \in E$  the set of {\em legal moves}, and finally, 
$r_i(e)$  is the cost of move $e  \in E$  for player  $i \in I$, 
called the {\em local  cost}. 

\subsection*{Strategies, plays, and effective costs}
\label{ss2a}  
A mapping  $s_i$  that  assigns a move  $(v, v')$  
to each position  $v \in V_i$  is a strategy of player  $i \in I$. 
(We restrict ourselves and all  players  
to their pure stationary strategies;   
no mixed or history dependent ones are considered in this paper.) 
Each {\em strategy profile}  $s = (s_1, \ldots, s_n)$  
uniquely defines a play  $p(s)$, that is, a walk in $G$  
that begins in the initial position  $s = v_0$  and 
goes in accordance with  $s$  in every position that appears. 
Obviously, $p(s)$  either terminates in  $t = v_t$  or cycles; 
respectively, it is called a terminal or a cyclic play. 
Indeed, after  $p(s)$  revisits a  position, it will 
repeat its previous moves thus making a ``lasso".  

The effective cost of $p(s)$  for a player  $i \in I$ is additive, that is,  
$$r_i(p(s)) = \sum_{e \in p(s)} r_i(e) \;\;\;  \text{if $p(s)$  is a terminal play;}$$    
$$r_i(p(s))= +\infty  \;\;\; \text{if  $p(s)$  is a cyclic play.}$$
In other word,  each player $i \in I$  pays the local cost  $r_i(e)$  
for every move $e \in p(s)$. 
Since a cyclic play  $p(s)$  never finishes and all local costs are positive, 
each player pays $+ \infty$.    
All players are minimizers.
Thus, a finite $n$-{\em person SP game} is defined. 
We study Nash-solvability (NS) of these games. 

\section{Nash equilibrium and Nash-solvability} 
\label{s3}  
Recall that a {\em strategy profile}  $s = (s_1, \dots, s_n)$  is called  
a {\em Nash equilibrium} (NE)  if 
$r_i(s') \geq  r_i(s)$  whenever  $s'$  differs from  $s$ 
only by the strategy of player  $i$, that is, 
$s_j  = s'_j$  for all  $j \neq  i$. 
In other words, no player  $i \in I$   can make a  profit 
by changing his/her strategy  provided  all other players 
keep  their strategies unchanged. 

\medskip

The Bi-SP conjecture means exactly that all finite two-person SP games 
(with positive local costs) are NS. 
Indeed, a pair of strategies  $s = (s_1, s_2)$  
realizes a bi-shortest path in $G$  if and only if  $s$ 
is a  NE in the corresponding two-person SP game. 

\medskip  

However, a three-person SP game, even with positive local costs, may be not NS; see    
\cite[Tables 2,3 and Figure 2]{GO14}.  

\bigskip

Digraph  $G = (V,E)$  is called {\em bidirected} if
each non-terminal  move in it is reversible, that is,
$(u,w) \in E$  if and only if  $(w,u) \in E$
unless  $u = t$ or $w = t$. 
We conjecture that every $n$-person SP game on a finite bidirected digraph is NS.      

\section{Essential properties of cost functions} 
\label{s4} 
\subsection*{$k$-total costs and rewards} 
SP games can be viewed as a very special class within the so-called 
finite deterministic stochastic games 
with perfect information with $k$-total effective reward \cite{BEGM17}. 
(Negated costs are called payoffs or rewards.) 
The limit mean payoff \cite{Gil57,LL69}, most common in the literature, and  
the total reward \cite{TV98,BEGM18}, 
correspond to  $k=0$  and  $k=1$, respectively \cite{BEGM17}.  
The family of $k$-total effective rewards is nested with respect to $k$, that is,  
$k$-total rewards can be properly embedded 
into $(k+1)$-total rewards \cite{BEGM17}.

Mostly, the two-person zero-sum case is studied in the literature.
Yet, all main concepts and definitions 
can be naturally extended to the $n$-person case; 
in particular, to the two-person but not necessarily zero-sum case.  
The obtained games may have no NE  already for $n=2$ and $k=0$; see \cite{Gur88}. 
Since they  are  $k$-nested, NS may fail for any $n \geq 2$  and  $k \geq 0$. 

Yet, NS becomes an open problem for $n = 2$  and  $k = 1$,   
provided we require that all local rewards are negative, 
or in other words, that all local  costs are positive \cite[Section 8]{BEGM17}.  
This is an alternative view at the Bi-SP conjecture. 

\subsection*{Positive costs and Gallai's Potential Transformation} 
The latter requirement: 

\begin{itemize}
\item[(i)] $\;\;\; r_i(e) > 0$  for each player $i \in I$ and directed edge $e  \in E$ 
\end{itemize}

\noindent 
can be replaced by a seemingly weaker 
(but in fact, equivalent) one:  

\begin{itemize}
\item[(ii)] $\;\;\; \sum_{e \in C} r_i(e) > 0$  for each player $i \in I$ and 
directed cycle  $C$  in  $G$. 
\end{itemize}

Implication  (i) $\Rightarrow$ (ii)  is  obvious. 
Conversely, if (ii) holds, one can enforce  (i)  
applying the following potential transformation \cite{Gal58}. 
Choose an arbitrary mapping  $x : V \rightarrow \RR$ 
and replace  $r_i(e)$  by  $r'_i(e) = r_i(e) + x(v) - x(v')$  
for every  $i \in I$ and $e = (v,v') \in E$. 
Obviously, this transformation does not change the game, since  
$r'(P) - r(P) = x(s) -  x(t) =  const$  
for every directed $(s,t)$-path  $P$. 
Furthermore,  $r'(C) = r(C)$  for  every directed cycle  $C$  in  $G$, and 
for  each  $r$  satisfying (ii) 
there exists a potential  $x$  such that (i)  holds for  $r'$  \cite{Gal58}. 

\section{Subgame perfect NE-free shortest path games}
\label{s5}
NE $s  = (s_1, \ldots, s_n)$  in a finite $n$-person SP game 
is called {\em uniform} if it is a NE with respect to
every initial position $s =  v_0 \in V \setminus  t$.
In the literature uniform NE (UNE) are frequently
referred to as {\em subgame perfect NE}.
By definition, any UNE is a NE, but not vice versa.
A large family of $n$-person UNE-free games 
can be found in \cite[Section 3.3]{GN21A} for  $n > 2$,
and even for  $n=2$  in \cite[the last examples in Figures 1 and 3]{BEGM12}. 
All these games have terminal payoffs, which is a special case the additive one.  
Hence, these games can be viewed as as a special case of the SP games.    

Every NE-free game contains a UNE-free subgame  \cite[Remark 3]{BGMOV18}.
Indeed, consider an arbitrary finite $n$-persoon NE-free SP game $\Gamma$
and  eliminate the initial position  $s =  v_0$  from its  graph $G$.
The  obtained subgame $\Gamma'$ is UNE-free.
Indeed, assume for contradiction that  $\Gamma'$  has a UNE  $s  = (s_1, \ldots, s_n)$.
Then, $\Gamma$  would also have a NE, which can be obtained by backward induction.
The player beginning in  $s = v_0$  chooses a move that maximizes his/her reward,
assuming that  $s$  is played in  $\Gamma'$  by all players.
Clearly, $s$  extended by this move forms a NE in $\Gamma$, which is a contradiction.

Thus, searching for a NE-free SP games
one should begin with a UNE-free SP game then  
trying to extend it with an acyclic prefix.
This was successfully realized in  \cite{GO14,BGMOV18}  for  $n=3$.
However, for  $n = 2$  all such tries failed.    

\subsection*{Acknowledgement}
The authors was partially supported by the RSF grant 20-11-20203.

\end{document}